\documentstyle[aps,multicol,epsf]{revtex}

\begin{document}


\title{
Language as an Evolving Word Web
}


\author{
S.N. Dorogovtsev,
$\!\!$\footnote{Departamento de F\'\i sica and Centro de F\'\i sica do Porto, Faculdade 
de Ci\^encias, 
Universidade do Porto, Rua do Campo Alegre 687, 4169-007 Porto, Portugal}
$\!\!$\footnote{A.F. Ioffe Physico-Technical Institute, 194021 St. Petersburg, Russia}
$\!\!$\footnote{E-mail: sdorogov@fc.up.pt and jfmendes@fc.up.pt}  
J.F.F. Mendes$^{\ast \ddag}$ 
}



\maketitle

\begin{abstract}
Human language can be described as a complex network of linked words. In such a treatment, each distinct word in language is a vertex of this web, and neighboring words in sentences are connected by edges. It was recently 
found 
\cite{fs01a} that the distribution of the numbers of connections of words in such a network 
is of a peculiar form which includes two pronounced power-law regions. 
Here we treat language as a self-organizing network of 
interacting words.
In the framework of this concept, we completely describe the 
observed Word Web structure without
fitting. 
\end{abstract}


\begin{multicols}{2}

$\phantom{.}$ 
\vspace{-12pt}

\narrowtext


How language evolves is 
a major challenge for linguistics and evolutionary biology \cite{ss97,d97,hsk98}
and 
an intriguing problem for 
other sciences \cite{s55,s57,nk99,npj00,n00c,nkn01}. The recent explosion of interest in 
networks \cite{ajb99,ha99,ba99,s01,w99},  
including the World Wide Web and Internet \cite{hppl98,lg98,lg99,ajb00}, biological networks \cite{jta00}, social \cite{ws98} and ecological webs \cite{wm00a}, networks of collaborations \cite{n01} etc., 
had the immediate consequence --- the treatment of human language as a complex network of distinct words \cite{fs01a}. 

This Word Web is arranged in the following way. The vertices of the web are 
the 
distinct words of language, and the undirected edges 
are connections between interacting words. It is not so easy to define the notion of word interaction in a unique way. Nevertheless, different reasonable definitions provide very similar structures of the Word Web. 
For instance, one can connect the nearest neighbors in sentences. 
Without going into detail, this means that the edge between two distinct words of language exists if these words are the nearest neighbors in at least one sentence in the bank of language. In such a definition, multiple links are absent. One also may connect the second nearest neighbors and account for other types of the correlations between words.  

Recently it was 
found
that this network has a complex architecture 
\cite{fs01a} which dramatically differs from classical random graphs extensively studied in the mathematical graph theory. In Ref. \cite{fs01a} 
the basic informative characteristic of the Word Web, the distribution of the numbers of connections of words, has been obtained empirically. 
In graph theory, the number of connections of a vertex is called its degree. 
The observed degree distribution of the Word Web has a long tail --- unlike the Poisson degree distribution for the classical random graphs. 
This indicates that the Word Web belongs to the same class as the World Wide Web and Internet \cite{ajb99,ha99}. 

Moreover, the degree distribution obtained in Ref. \cite{fs01a} has a 
complex form. It consists of two power-law parts with different exponents. This hampers any treatment but, on the other hand, makes possible to 
find an explanation of the basic structure of the 
word 
web in the framework of a general concept. Indeed, if one proposes a model which, 
without 
fitting, describes the 
empirical
degree distribution 
and reproduces the values of all the characteristic scales, 
the announced aim will be achieved (it is 
hardly possible 
to describe 
such a complex form perfectly by coincidence).  
Here we present the solution of this problem.   

Human language is certainly an evolving system. Its present structure is determined by its past evolution. 
This system is so complex that it can not be controlled but rather 
organizes itself while growing. 
We treat language as a growing network of 
interacting 
words. 
At its birth, a new word already 
interacts 
with several old ones.  
New 
interactions 
between old words emerge from time to time, and new edges arise.   

How do words find their collaborators in language? Here we use the idea of preferential linking (preferential attachment of new edges to vertices with higher numbers of connections) \cite{ba99}. This fruitful idea is a particular realization of the general concept of Simon \cite{s55,s57}. 
The simplest linear form of the preferential linking provides the power-law degree distributions for nets in which the average number of connections per vertex (the average degree) does not change during the growth \cite{ba99,dms00,krl00}.  
If the total number of connections increases faster than the number of vertices, and the average degree grows, the exponent of the degree distribution takes a 
different value \cite{dm01a}. For the explanation of the resulting structure of the Word Web, we combine these two processes of the edge emergence. 


We use the following rules of the network growth (see Fig. 1). At each time step, a new vertex (word) is added to the network, and the total number of vertices, $t$, plays the role of time. At 
its birth, the new word connects to several old ones. We do not know the original number of connections. We only know that it is of the order of $1$. It would be unfair to play with an unknown parameter to fit the experimental data, so we set this number to $1$ \cite{remark1}. We use the simplest natural version of the preferential linking, so a new word is connected to some old one $i$ with the probability proportional to its degree $k_i$, like in the Barab\'asi-Albert's model \cite{ba99}. In addition, at each increment of time, $ct$ new edges emerge between old words, where $c$ is a constant coefficient that characterises a particular network. 
The linear dependence appears if each vertex makes new connections with a constant rate, and we choose it as the most simple and natural. These new edges emerge between old words $i$ and $j$ with the probability proportional to the product of their degrees 
$k_i k_j$ \cite{ab00a,dm00a,remark2}. 

Two slightly different methods were used in Ref. \cite{fs01a} to construct 
the Word Web. 
The two resulting webs 
obtained after processing $3/4$ million words of the British National Corpus 
(a collection of text samples of both spoken and written modern British English)  
have nearly the same degree distributions, and each one contains about $470,000$ vertices. The average number of connections of a word (the average degree) is $\overline{k}\approx 72$. These are the only parameters of the Word Web we know 
and can use in the model. 

This stochastic model can be solved exactly but here, for a simple presentation, we use the continuous approximation. 
Such an approach was proved to describe quite well the degree distributions of networks growing under the mechanism of preferential linking \cite{ba99,dm01a,dm00a}. 
In our case, it provides the nonstationary degree distribution $P(k,t)$ very close to the exact one everywhere except of the narrow region $k<10$. One should emphasize that the continuous approach yields the exact values of the exponents of the distribution.   

In the continuous approximation, 
the degrees of the vertices born at time $s$ and observed at time $t$ are substituted by their average value $k(s,t)$. For the large network, the evolution of $k(s,t)$ is described by the simple equation    

\begin{equation}
\frac{\partial k(s,t)}{\partial t} = 
(1+2ct)\frac{k(s,t)}{\int^t_0du\,k(u,t)}
\,   
\label{1}
\end{equation}   
with the obvious boundary condition $k(t,t)=1$. 
The nature of this equation can be easily understood. 
The ratio on the right hand side is a direct consequence of the preferential attachement. At each time step, $1 + 2ct$ ends of new edges are distributed preferentially. 
Indeed, one such an end belongs to the edge coming from a new word, and the 
others 
are the ends of the $ct$ new edges emerging between old words.
Here, we have presented heuristic arguments but Eq. \ref{1} can be derived more 
strictly \cite{dm00a}. 

One sees that the total degree of the network is 
$\int^t_0du\,k(u,t) = 2t + ct^2$, so its average degree is equal to 
$\overline{k}(t) = 2 + ct$. The present value of the average degree of the Word Web is close to $72$, hence $1 \ll ct \approx 70$.
The solution of Eq. \ref{1} is of a singular form

\begin{equation}
k(s,t) = 
\left(\frac{ct}{cs}\right)^{1/2} \left(\frac{2+ct}{2+cs}\right)^{3/2}
\,   
\label{2}
\end{equation}  
which indicates the presence of two distinct regimes in this problem. 
From Eq. \ref{2},  
we immediately obtain the nonstationary degree distribution \cite{remark3}

\begin{equation}
P(k,t) = \frac{1}{ct} \frac{cs(2+cs)}{1+2cs} \frac{1}{k}
\, ,  
\label{3}
\end{equation}  
where $s=s(k,t)$ is the solution of Eq. \ref{2}. 

One sees from Eqs. \ref{2} and \ref{3} that this nonstationary distribution has two regions with different behaviors separated by the crossover point 
$k_{cross} \approx \sqrt{ct}(2+ct)^{3/2}$. The crossover moves in the direction of large degrees while the network grows. 
Below this point, the degree distribution is stationary, 
$P(k) \cong \frac{1}{2}k^{-3/2}$ (we use the fact that in the Word Web $ct \gg 1$). Above the crossover point, we obtain the behavior  
$P(k,t) \cong \frac{1}{4}(ct)^3 k^{-3}$, so that the degree distribution is nonstationary in this region. Thus, the model provides two distinct values for the degree distribution exponent, $3/2$ and $3$. 

The degree distribution has one more important characteristic point, the cutoff 
produced by the size effect. 
Its position $k_{cut}$ is easily estimated from the condition that one vertex in the network is of degree exceeding $k_{cut}$, that is   
$t\int_{k_{cut}}^\infty dk\,P(k) \sim 1$ 
and thus 
$k_{cut} \sim \sqrt{t/8}(ct)^{3/2}$. 
Here we do not present the complete exact result which can be obtained using the 
master equation approach. 
The infinite limit of the exact degree distribution takes the simple form 
$P(k,t \to \infty) = \frac{1}{2}B(k,3/2)$ where $B(\,,\,)$ is the beta-function. Minor deviations from the continuous approximation are visible only for $k<10$. 

In Fig. 2, we plot the degree distribution of the model 
(the solid line). To obtain the theoretical curve, we used Eqs. \ref{2} and \ref{3} with the known parameters of the Word Web. 
The deviations from the continuous approximation are accounted for in the small $k$ region, $k<10$. One sees that the agreement with the empirical data \cite{fs01a} is excellent. Note that we do not use any fitting. 
For a better comparison, in Fig. \ref{f2}, the theoretical curve is displaced upward 
(we have to exclude two experimental points with the smallest $k$ since these points are dependent on the method of the construction of the Word Web, and any comparison in this region is meaningless in principle). 

\begin{figure}
\epsfxsize=90mm
\epsffile{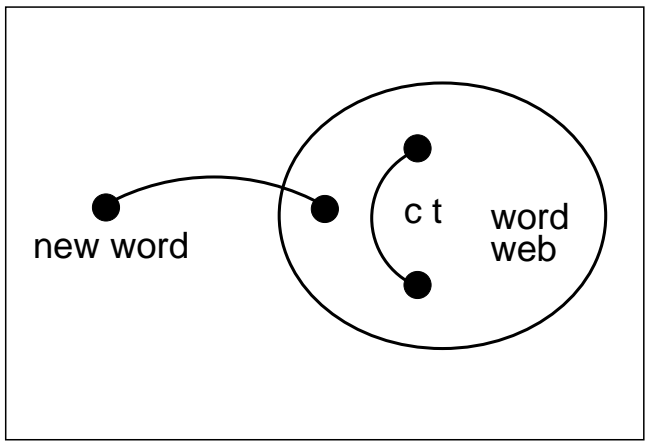}
\caption{
Scheme of the Word Web growth. 
At each time step a new word is appear, so $t$ is the total number of words. It connects to some preferentially chosen old word. Simultaneously, $ct$ new edges emerge between pairs of preferentially chosen old words. 
All the edges are undirected. 
We use the simplest kind of the preferential attachment when a node is chosen with the probability proportional to the number of its connections.  
}
\label{f1}
\end{figure}

\begin{figure}
\epsfxsize=90mm
\epsffile{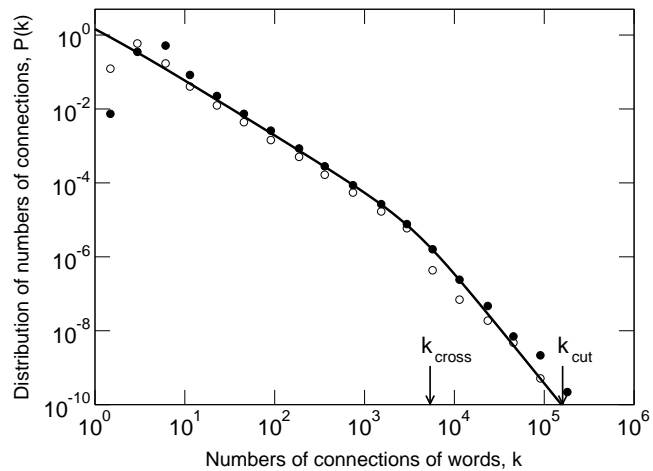}
\caption{
The distribution of the numbers of connections (degrees) of words in the Word Web in a log-log scale. The solid line is the result of our calculation using the parameters of the Word Web, the size $t \approx 470,000$ and the average number of connections $\overline{k}(t) \approx 72$. 
Empty and filled circles show the distributions of the numbers of 
connections obtained in Ref. \protect\cite{fs01a} for the two different methods of construction of the Word Web.   
In the region $k<10$, where the deviations of the continuous approximation from the exact solution of the model are noticeable, we present the exact solution. 
The arrows indicate the theoretically obtained 
point of crossover, $k_{cross}$, between the regions with the exponents $3/2$ and $3$ and the cutoff $k_{cut}$ of the power-law dependence due to the size effect.  
For a better comparison, the theoretical curve is displaced upward (note that the comparison is impossible in the region of the smallest $k$ where the experimentally obtained 
distribution essentially depends on the definition of the Word Web). 
}
\label{f2}
\end{figure}

From the relations obtained above, we find the characteristic values for the crossover and cutoff, $k_{cross} \approx 5.1\times 10^3$, that is, $\log_{10} k_{cross} \approx 3.7$, and $\log_{10} k_{cut} \approx 5.2$. 
From Fig. \ref{f1}, one sees that these values coincide with the experimental ones. 
As far as we know, it is the first time that such complex empirically obtained data for networks are described without fitting. 
We should emphasize that the extent of agreement is truly surprising. The minimal model does not account for numerous, at first sight, important factors, e.g., the death of words, the variations of words during the evolution of language, 
etc. \cite{remark4}.  
The agreement is convincing since it is approached over the whole range of values of $k$, 
that is, 
over five decades. 
In fact, the Word Web turns out to be very convenient in this respect since the total number of edges in it is extremely high, about $3.4\times10^7$ edges, and the value of the cutoff degree 
is large. 


Note that few words are in the region above the crossover point $k_{cross} \approx 5.1\times 10^3$. With the growth of language, $k_{cross}$ increases rapidly but, as it follows from our relations, the total number of words of degree greater than $k_{cross}$ does not change. It is a constant of the order of 
$1/(8c) \approx t/(8\overline{k}) \sim 10^3$, that is of the order of the size of 
a small set of words forming the 
kernel lexicon of the British English which was estimated as $5,000$ words \cite{fs00a} and is the most important core part of language. Therefore, our concept suggests that the number of words in this part of language does not depend essentially of its size. 
Formally speaking, 
this is determined by the value of the average rate $c$ with which words find new partners in language.  

There exist many obvious ways to improve the 
minimal model used above. 
Nevertheless, at present, such attempts seem  
rather meaningless since, 
as we have noted, it is hard to define 
rigorously the procedure of 
the Word Web construction, 
and the experimental data do not allow us to make a better comparison.  

We have proposed a simple stochastic theory of evolution of human language based on 
the treatment of language as an evolving network of 
interacting 
words. 
The structure of language is the 
result of the self-organization of the Word Web during its growth.  
The key result is the distribution of numbers of connections of words. 
We have found that the self-organization produces the most connected small kernel lexicon of language, 
size of which does not change essentially along the language evolution. The degree distribution of words in this core of language crucially differs from the degree 
distribution of the rest. 
We have shown that the basic characteristic of the Word Web structure, namely the degree distribution,  
does not depend on the rules of language but is determined by the general principles of the evolutionary dynamics of the Word Web. 
We would like to note that the successful description is important since the 
recent progress in the understanding of numerous stochastic multiplicative processes in Nature is based on the famous Simon's model \cite{s55,s57} 
which was 
originally applied to the description of the structure of
human language. 



\end{multicols}


\begin{references} 


\bibitem{fs01a} R. Ferrer, R. V. Sol\'e, 
{\em Proc. Royal Soc. London B}, submitted, 
{\em Santa Fe Working Papers} 01-03-016 (2001), available at http://www.santafe.edu/sfi/publications/Abstracts/01-03-016abs.html. 

\bibitem{ss97} J. M. Smith, E. Sz\"athm\'ary, {\em The Major Transitions in Evolution} 
(Oxford University Press, Oxford, 1997). 

\bibitem{d97} T. W. Deacon, {\em The Symbolic Species: The Coevolution of Language and the Brain} (W. W. Norton \& Co, New York, 1997). 

\bibitem{hsk98} J. R. Hurford, M. Studdert-Kennedy, C. Knight, Eds., 
{\em Approaches to the Evolution of Language} (Cambridge University Press, Cambridge, 1998). 

\bibitem{s55} H. A. Simon, {\em Biometrica} {\bf 42}, 425 (1955).

\bibitem{s57} H. A. Simon (1957), {\em Models of Man} (Wiley, New York, 1957). 

\bibitem{nk99} M. A. Nowak and D. C. Krakauer, {\em Proc. Natl. Acad. Sci. U. S. A.} {\bf 96}, 
8028 (1999). 

\bibitem{npj00} M. A. Nowak, J. B. Plotkin, V. A. Jansen, {\em Nature}  {\bf 404}, 
495 (2000). 

\bibitem{n00c} M. A. Nowak, {\em J. Theor. Biology} {\bf 204}, 179 (2000).  

\bibitem{nkn01} M. A. Nowak, N. L. Komarova, P. Niyogi, 
{\em Science} {\bf 404}, 180 (2001).

\bibitem{ajb99}  R. Albert, H. Jeong, A.-L. Barab\'{a}si, {\em Nature} {\bf 401}, 
130 (1999). 

\bibitem{ha99} B. A. Huberman, L. A. Adamic, {\em Nature} {\bf 401}, 
131 (1999). 

\bibitem{ba99} A.-L. Barab\'{a}si, R. Albert, {\em Science} {\bf 286}, 509 (1999). 

\bibitem{s01}  S. H.  Strogatz, {\em Nature} (London) {\bf 410}, 
268 (2001). 

\bibitem{w99} D. J. Watts, {\em Small Worlds: The Dynamics of Networks between Order and Randomness} 
(Princeton University Press, Princeton, 1999). 
 
\bibitem{hppl98} B. A. Huberman, P. L. T. Pirollo, J. E. Pitkow, R. M. Lukose, 
{\em Science} {\bf 280}, 95 (2000). 

\bibitem{lg98} S. Lawrence, C. L. Giles, {\em Science} {\bf 280}, 98 (1998).  

\bibitem{lg99} S. Lawrence, C. L. Giles, {\em Nature} {\bf 400}, 107 (1999).  

\bibitem{ajb00} R. Albert, H. Jeong, A.-L. Barab\'asi,
{\em Nature} {\bf 406}, 378 (2000).
 
\bibitem{jta00} H. Jeong, B. Tombor, R. Albert, Z. N. Oltwai, A.-L. Barab\'{a}si,  
{\em Nature} {\bf 407}, 651 (2000).  

\bibitem{ws98} D. J. Watts, S. H. Strogatz, {\em Nature} {\bf 393}, 440 (1998). 

\bibitem{wm00a} R. J. Williams and N. D. Martinez, 
{\em Nature} {\bf 404}, 180 (2000).

\bibitem{n01} M. E. J. Newman, {\em Proc. Natl. Acad. Sci. U. S. A.} {\bf 98}, 
404 (2001).  

\bibitem{dms00} S. N. Dorogovtsev, J. F. F. Mendes, A. N. Samukhin,  
{\em Phys. Rev. Lett.} {\bf 85}, 4637 (2000). 

\bibitem{krl00} P. L. Krapivsky, S. Redner, F. Leyvraz, 
{\em Phys. Rev. Lett.} {\bf 85}, 4633 (2000). 

\bibitem{dm01a} S. N. Dorogovtsev, J. F. F. Mendes, 
{\em Phys. Rev. E} {\bf 63}, 025101 (R) (2001). 

\bibitem{remark1} One can check that the introduction of this parameter does not change noticeably the degree distribution of the Word Web. 

\bibitem{ab00a} R. Albert, A.-L. Barab\'{a}si, 
{\em Phys. Rev. Lett.} {\bf 85}, 5234 (2000). 


\bibitem{dm00a} S. N. Dorogovtsev, J. F. F. Mendes, {\em Europhys. Lett.} {\bf 52}, 33 (2000). 

\bibitem{remark2} A similar model was recently applied to 
the description of networks of  
coauthorships in the scientific literature 
[A.-L. Barab\'{a}si, H. Jeong, Z. N\'eda, E. Ravasz, 
A. Schubert, T. Vicsek, {\em arXiv cond-mat,} 0104162 (2001), 
available at http://arXiv.org/abs/cond-mat/0104162]. 






\bibitem{remark3} Here we use the standard expression valid in the continuous approach, 
$P(k,t) = -[t\partial k(s,t)/\partial s]^{-1} \left.\right|_{s=s(k,t)}$. 

\bibitem{remark4} The clustering coefficient of the Word Web takes large values \protect\cite{fs01a}. One can easily explain this by incorporating 
the following features. (i) A word simultaneously makes not one but several connections. (ii) These edges usually come to already interacting words. 
Here we do not account for these circumstances  
since we are interested only in the degree distribution. 

\bibitem{fs00a} R. Ferrer, R. V. Sol\'e, 
{\em J. Quantitative Linguistics}, in press, 
{\em Santa Fe Working Papers} 00-12-068 (2000), available at   
http://www.santafe.edu/sfi/publications/Abstracts/00-12-068abs.html.

\bibitem{thanks} 
We thank A. N. Samukhin for helpful discussions and G. Tripathy for thorough reading our manuscript. Supported by the project POCTI/1999/FIS/33141.


 






















 






\end{references}
\end{document}